\documentclass[prl,twocolumn,showpacs,superscriptaddress,aps]{revtex4}
\usepackage{amsfonts}
\usepackage{amsmath}
\usepackage{amssymb}
\usepackage{graphicx,color}
\usepackage{dcolumn}
\usepackage{times}

\begin{document}

\title{Three-component soliton states
in spinor $F=1$ Bose-Einstein condensates}

\author{T.~M.~Bersano}
\affiliation{Washington State University,
Department of Physics \& Astronomy,
Pullman, WA 99164 USA}

\author{V.~Gokhroo}
\affiliation{Washington State University,
Department of Physics \& Astronomy,
Pullman, WA 99164 USA}

\author{M.~A.~Khamehchi}
\affiliation{Washington State University,
Department of Physics \& Astronomy,
Pullman, WA 99164 USA}

\author{J. D'Ambroise}
\affiliation{ Department of Mathematics, Computer \& Information Science, State University of New York (SUNY) College at Old Westbury, Westbury, NY, 11568, USA}

\author{D. J.\ Frantzeskakis}
\affiliation{Department of Physics, National and Kapodistrian University of Athens, Panepistimiopolis, Zografos, Athens 15784, Greece}

\author{P. Engels}
\affiliation{Washington State University,
Department of Physics \& Astronomy,
Pullman, WA 99164 USA}

\author{P. G. Kevrekidis}
\affiliation{Department of Mathematics and Statistics, University of Massachusetts,
Amherst, MA, 01003, USA}

\begin{abstract}

  Dilute-gas Bose-Einstein condensates are an exceptionally versatile testbed for the investigation of novel solitonic structures. While matter-wave solitons in one- and two-component systems have been the focus of intense research efforts, an extension to three components has never been attempted in experiments,
  to the best of our knowledge. Here, we experimentally demonstrate the existence of robust dark-bright-bright (DBB) 
and dark-dark-bright (DDB) solitons in a spinor $F=1$ condensate. We observe lifetimes on the order of  hundreds of milliseconds for these structures. Our theoretical analysis, 
based on a multiscale expansion method, shows that small-amplitude solitons of these types 
obey universal long-short wave resonant interaction models, namely Yajima-Oikawa systems. 
Our experimental and analytical findings are corroborated by direct 
numerical simulations highlighting the persistence of, e.g., the DBB states, as well 
as their robust oscillations in the trap.

\end{abstract}

\pacs{03.75.Mn, 03.75.Lm}

\maketitle

Solitons are localized waves propagating undistorted in nonlinear
dispersive media. They play a key role in numerous physical contexts \cite{dp}. 
Among the various systems that support solitons, dilute-gas Bose-Einstein condensates 
(BECs) \cite{pita,rom} provide a particularly versatile testbed for the investigation of solitonic 
structures~\cite{emergent,fkh,djf}. In single-component BECs, solitons have been 
observed either as robust localized pulses (bright solitons)~\cite{b1,b2,b3,b4,b5} 
or density dips in a background matter wave 
(dark solitons)~\cite{han1,nist,dutton,bpa,hamburg,hambcol,technion,kip,kip2,engels},
typically in BECs with attractive or repulsive interatomic interactions, respectively.
Extending such studies to two-component BECs has led to rich additional dynamics.
Solitons have been observed in binary mixtures of different spin states of the same atomic species,
so-called pseudo-spinor BECs~\cite{Hall1998a,chap01:stamp}. In particular,
dark-bright (DB) \cite{hambdb,pe1,pe2,pe3,azu}, and related SO$(2)$
rotated states in the form of dark-dark solitons \cite{pe4,pe5},
have experimentally been created in binary $^{87}$Rb BECs.
Interestingly, although such BEC mixtures feature repulsive intra- and
inter-component interactions, bright solitons do emerge
due to an effective potential well created by the dark soliton through the
inter-component interaction \cite{small}.
Such mixed soliton states have been proposed for potential applications.
Indeed, in the context of optics where these structures were pioneered~\cite{seg1,seg2},
the dark soliton component was proposed to act as an adjustable waveguide for weak
bright solitons \cite{kivshar}. In multicomponent BECs, compound solitons of the
mixed type could also be used for all-matter-wave waveguiding, with the dark soliton
building an effective conduit for the bright one, similar to all-optical
waveguiding in optics \cite{bld}. 
Apart from pseudo-spinor BECs, such 
mixed soliton states have also been predicted to occur in genuinely spinorial BECs,  
composed of different Zeeman sub-levels of the same 
hyperfine state~\cite{Stenger1998a,kawueda,stampueda}. Indeed, 
pertinent works \cite{bdspinor,chin}
have studied the existence and dynamics of 
DB soliton complexes in spinor $F=1$ BECs. 
However, experimental observation of such states has not been reported so far.

Here we report on the systematic experimental generation of three-component DB soliton complexes,
of the dark-bright-bright (DBB) and dark-dark-bright (DDB) types, in a spinor $F=1$ 
condensate of $^{87}$Rb atoms. While DB solitons normally consist of two 
atomic states (e.g., two $F=1$ Zeeman sublevels or a combination of Zeeman sublevels
of $F=1$ and $F=2$ states of $^{87}\text{Rb}$ \cite{pe1,pe2,pe3,azu,pe4,pe5}), here
we use all three Zeeman $F=1$ sublevels to generate three-component solitons in an
elongated atomic cloud. In our theoretical analysis, we employ a multiscale expansion method to
derive such vector soliton solutions of the pertinent
Gross-Pitaevskii equations (GPEs). We thus show that DBB and DDB solitons
can be approximated by solutions of 
 Yajima-Oikawa systems~\cite{yajima,magnon,myo}. We thus provide a
connection with universal long-short wave resonant interaction (LSRI)
processes \cite{vz} which appear in a wide range of contexts, including
plasmas \cite{yajima}, condensed matter \cite{magnon}, hydrodynamics \cite{hydro},
nonlinear optics \cite{nlo}, negative refractive index media \cite{mm}, etc.
Our experimental and analytical identification of these spinor
solitonic structures is corroborated by direct numerical simulations.

To begin our discussion of the three-component solitonic structures,
we first present examples for their realization in experiments.
The three components are given by the three different Zeeman sublevels of the $F=1$ state of $^{87}$Rb,
and are designated by their magnetic quantum numbers
$|F,m_F\rangle = |1,-1\rangle,~|1,0\rangle$ and $|1,+1\rangle$.
The experiments begin with a single-component BEC of approximately $0.8\times 10^6$~ atoms.
The atoms are confined in an elongated harmonic trap with 
frequencies $\{\omega_x,\omega_y,\omega_z\}=2\pi \times\{1.4,176,174\}$~Hz, where
$z$ is the vertical direction.
The trap is formed by a focused dipole laser beam and is independent of the atomic hyperfine state.
A magnetic bias field of $45.5$~G is applied along the weakly confining direction. This field
leads to a Zeeman splitting of the energy levels. As a consequence, populations
can be transferred between the three states by using radio frequency (RF) pulses
or adiabatic radio frequency sweeps.

\begin{figure}
\begin{center}
\includegraphics[width=3in]{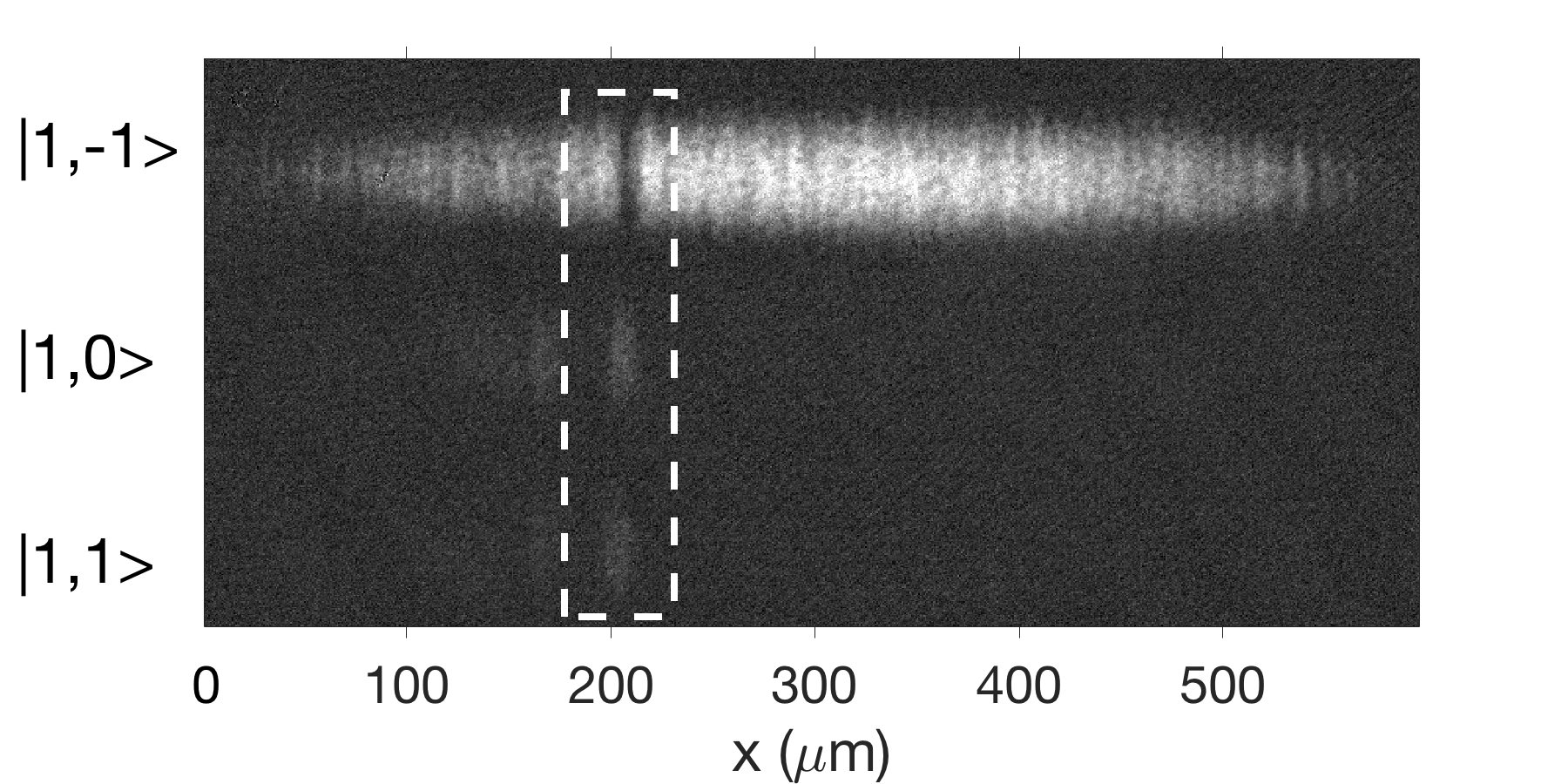}
\end{center}
\hspace{-0.3in}
\includegraphics[width=2.8in]{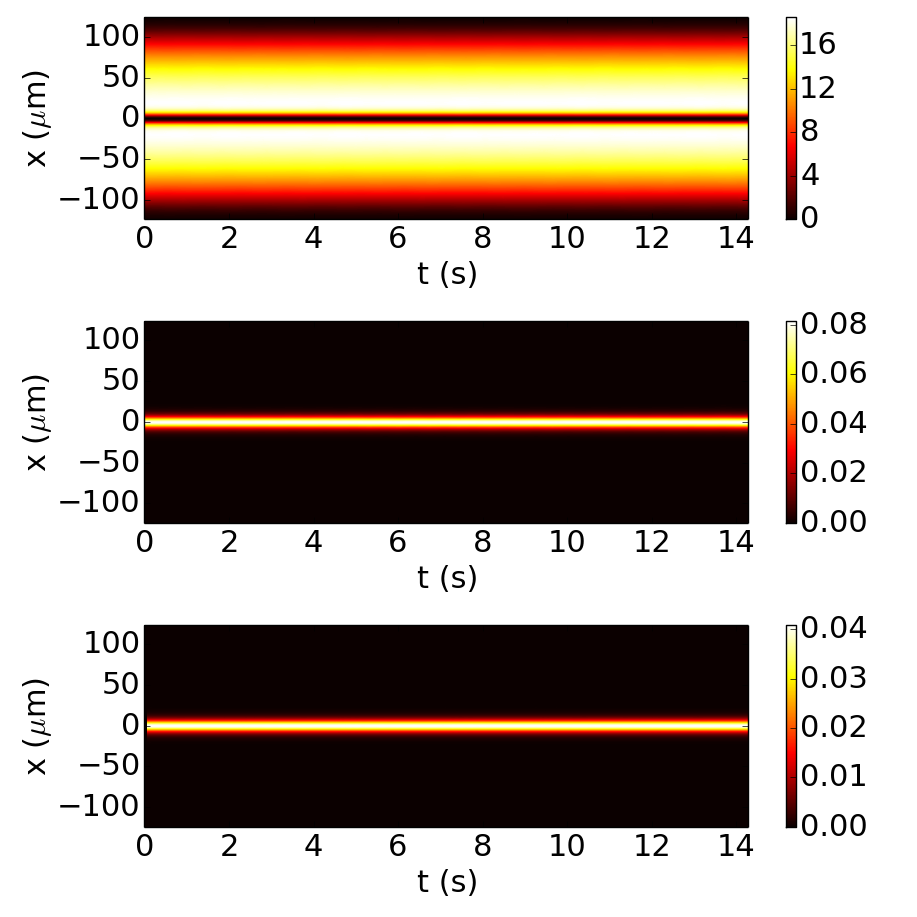}
\caption{(Color online) Top panel: Experimental ToF image of a DBB soliton (boxed). In each case, to verify the stability of the soliton a $100~\text{ms}$ in-trap evolution time is applied after DBB soliton formation. The relative population of the three states
  $|1,-1\rangle$, $|1,0\rangle$ and $|1,1\rangle$ is 97:2:1.
  Bottom panels: Numerical simulation of the evolution of
  the densities $|\psi_{-1}|^2$ (top), $|\psi_{0}|^2$ (middle), and $|\psi_{+1}|^2$ (bottom) as per
  Eqs.~(\ref{gp1})-(\ref{gp2}). The DBB soliton is found to be extremely
  robust.}
\label{DBB_soliton}
\end{figure}

To generate DBB solitons such as the ones shown in the top panel of Fig.~\ref{DBB_soliton},
we begin with all atoms in the $|1,-1\rangle$ state. A small fraction of
atoms is transferred to the $|1,0\rangle$ state using a RF sweep.  
Subsequently, a weak magnetic gradient is applied along the long axis of the BEC 
(i.e., the x-axis) for approximately $2-3$~sec. 
Since the states have different magnetic moments, this induces superfluid-superfluid
counterflow and leads to the formation of DB solitons; see details of this technique
in Refs.~\cite{pe1,pe4}. In the present experiment, the dark solitons reside in
the $|1,-1\rangle$ state and the bright component is formed
by the $|1,0\rangle$ state. After the removal of the gradient, a second RF transfer
moves a fraction of the atoms from the $|1,0\rangle$ state to the $|1,+1\rangle$
state, forming a DBB soliton. After a variable evolution time during 
which the solitons are kept in the trap, a Stern-Gerlach imaging technique is 
used to individually image all three components in one single run of the experiment 
\cite{imaging}.

To theoretically trace the formation of such compound soliton structures,
we resort to mean-field theory. In this framework, the wave functions
$\psi_{\pm1,0}(x,t)$ of the three hyperfine components
($m_F=\pm1,0$) of a quasi one-dimensional $F=1$ spinor BEC obey the following 
GPEs \cite{mart,yuri,bdspinor,chin}:
\begin{subequations}
\begin{eqnarray}
i \partial_t \psi_{\pm 1} &=& \mathcal{L}\psi_{\pm 1} +
\lambda_a\left(|\psi_{\pm 1}|^2+|\psi_{0}|^2-|\psi_{\mp 1}|^2\right)\psi_{\pm 1}
\nonumber \\
&+& \lambda_a\psi_0^2\bar{\psi}_{\mp 1},
\label{gp1} \\[1.0ex]
i \partial_t \psi_{0} &=& \mathcal{L}\psi_{0} +
\lambda_a \left(|\psi_{1}|^2+|\psi_{-1}|^2\right)\psi_{0}
\nonumber \\
&+&
2\lambda_a\psi_{-1}\bar{\psi}_0 \psi_{+1},
\label{gp2}
\end{eqnarray}
\end{subequations}
where $\mathcal{L} = -\frac{1}{2}\partial_x^2
+ V(x) + \lambda_s\left( |\psi_{-1}|^2+|\psi_{0}|^2+|\psi_{1}|^2\right)$ and
$V(x) = (1/2)\Omega^2 x^2$, with $\Omega = \omega_x/\omega_\perp$. We use
$\omega_x=1.4$~Hz and $\omega_\perp=175$~Hz, as per the experimental set up.
Finally, the coupling coefficients for ``symmetric''
spin-independent and ``antisymmetric'' spin-dependent interaction terms are given by
$\lambda_s=\frac{2}{3}\left(a_0 + 2a_2\right)/a_\perp$ and $\lambda_a=\frac{2}{3}\left( a_2 - a_0\right)/a_\perp$, respectively, where $a_0$ and $a_2$ correspond to
s-wave scattering lengths of two atoms in the scattering channels with total spin 0 and 2, and
$a_\perp=\sqrt{\hbar/(M \omega_\perp)}$, with $M$ being the atomic mass of Rb. In our case,
$\lambda_s \approx 5.2 \times 10^{-3}$ and $\lambda_a \approx -2.4 \times 10^{-5}$, i.e.,
$\lambda_a/|\lambda_s|$ is a small parameter. 

Based on this fact, it can readily be observed that 
--in the absence of the trap, and ignoring the spin-dependent
interactions-- the GPEs~(\ref{gp1})-(\ref{gp2}) reduce to the
completely integrable Manakov system \cite{manakov}. This model admits vector soliton
solutions of the mixed type (i.e., DB soliton complexes --cf., e.g., Ref.~\cite{feng})
which may persist in the presence of the spin-dependent terms of
Eqs.~(\ref{gp1})-(\ref{gp2}) and the trap. Nevertheless, to better understand the
role of the spin-dependent nonlinear interatomic interactions, we employ (for $V(x)=0$)
a multiscale expansion method to find approximate DBB soliton solutions 
of Eqs.~(\ref{gp1})-(\ref{gp2}). The lines of analysis are similar to those 
of Ref.~\cite{bdspinor}, but with an important difference: 
in Ref.~\cite{bdspinor}, a single-mode approximation (SMA) \cite{pu,yuri} was used 
for the ``symmetric'' states $|1,-1\rangle$ and $|1,+1\rangle$; here, 
instead, we rely on the smallness of $\lambda_a/|\lambda_s|$ 
to identify different types of solitons (dark and bright) for the states $|1,-1\rangle$ 
and $|1,+1\rangle$. Then, upon imposing nontrivial boundary conditions for the 
$|1,-1\rangle$ component and trivial ones for the $|1,0\rangle$ and $|1,+1\rangle$ components, 
we can derive DBB soliton solutions for sublevels $m_{F}=-1,0,+1$, respectively.
More specifically, we can show that the wave functions of DBB solitons take the following form:
\begin{equation}
\begin{array}{rcl}
\!\!\!\!\!\!\!\!\!\!\!\!
\psi_{-1}&\approx &\sqrt{n_0 +\epsilon n(X,T)}
e^{-i\mu_{-1} t + i\epsilon^{1/2}C^{-1}(\lambda_s+\lambda_a)\int n dX},
\\[1.0ex]
\psi_{0,+1}&\approx&\epsilon^{3/4} q_{0,+1}(X,T)
e^{i\left[Cx-\left(\frac{1}{2}C^2 +\mu_{0,+1}\right) t\right]},
\label{ps2}
\end{array}
\end{equation}
where $\epsilon$ is a formal small parameter, $\mu_{\mp 1}=(\lambda_s \pm \lambda_a)n_0$
and $\mu_0 = \lambda_s n_0$ are the chemical potentials of the three components, while $n_0$ and
$C^2 = \mu_{-1}$ denote, respectively, the steady-state density and the speed of sound of the $|1,-1\rangle$
component. Finally, the unknown functions $n(X,T)$ and $q_{0,+1}(X,T)$, which depend 
on the stretched variables $X=\epsilon^{1/2}(x-Ct)$ and $T=\epsilon t$, obey the system:
\begin{equation}
\begin{array}{rcl}
&&\partial_{T}n =-\sqrt{\mu_{-1}}\partial_{X}\left(|q_0|^{2}\right)
-\sqrt{\mu_{+1}}\partial_{X}\left(|q_{+1}|^{2}\right),
\\[1.0ex]
&&i\partial_{T}q_0+\frac{1}{2}\partial_{X}^{2}q_0-\lambda_s nq_0 =0,
\\[1.0ex]
&&i\partial_{T}q_{+1}+\frac{1}{2}\partial_{X}^{2}q_{+1}-(\lambda_s-\lambda_a) nq_{+1}=0.
\label{myo}
\end{array}
\end{equation}
The above system, which models long-short-wave resonance (LSRI) interaction \cite{vz},
is the multicomponent generalization of the so-called Yajima-Oikawa (YO) system, 
originally derived to describe the interaction of Langmuir and sound waves 
in plasmas~\cite{yajima}. In fact, Eqs.~(\ref{myo}) constitute the so-called 
multicomponent YO (mYO) system, originally introduced in the context of many-component 
magnon-phonon systems \cite{magnon}, which generalizes the YO model \cite{myo}. 
This model has recently attracted considerable attention due to its variety of solutions 
and interesting soliton collision properties  \cite{kanna1,kanna2,mar1}. Similarly to 
the single-component YO model, the mYO system is completely integrable, and possesses 
soliton solutions of the form \cite{kanna1}: 
$n \propto -{\rm sech}^2(K_s X-\Omega_s T)$ and $q_{0,+1} \propto {\rm sech}(K_s X-\Omega_s T)$,
where $K_s,~\Omega_s$ are constants. When substituted into Eqs.~(\ref{ps2}),
these expressions give rise to approximate DBB solitons, for the $m_{F}=-1,0,+1$ spin components, respectively.

\begin{figure}[tbp]
\begin{center}
\includegraphics[width=3in]{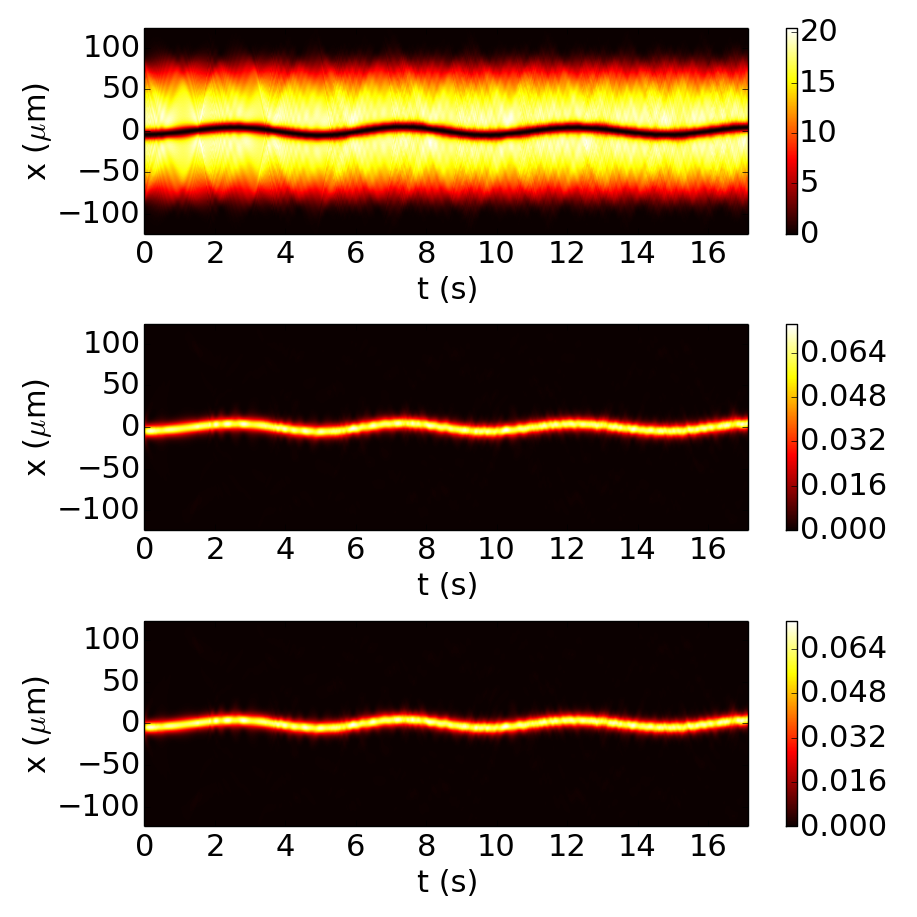}
\end{center}
\hspace{-0.3in}
\caption{(Color online) Top panel:  Numerical simulation of the evolution of
  the densities $|\psi_{-1}|^2$ (top), $|\psi_{0}|^2$ (middle), and $|\psi_{+1}|^2$ (bottom) as per
  Eqs.~(\ref{gp1})-(\ref{gp2}) in the case where a small spatial displacement of the DBB structure is initiated at $t=0$. It is observed that robust
oscillations of the DBB structure persist for several seconds.}

\label{osc}
\end{figure}

Direct numerical simulations corroborate our experimental and
analytical findings. In the performed simulations, the total number of atoms and energy of the system
(cf., e.g., Ref.~\cite{yuri} for definitions) are conserved up to a
negligible error. First, we have found (results not shown here) that 
the above mentioned small-amplitude solitons persist for large amplitudes. Second, 
apart from the traveling DBB solitons, we were also able to identify robust stationary 
such structures in the presence of the trap. A pertinent example is shown in the three 
bottom panels of Fig.~\ref{DBB_soliton}, where a robust DBB soliton, persisting for 
long time, is shown. This solution is constructed by identifying, 
at first, a stationary DB soliton state of the form
$\psi_{-1}(x,0)=\sqrt{\lambda_s^{-1}[\mu_{-1}-V(x)]}\tanh(\sqrt{\mu_{-1}}x)$,
$\psi_0(x,0)=A{\rm sech}(\sqrt{\mu_0} x)$ and $\psi_{+1}(x,0)=0$. Then,
in line with our experimental protocol, switching on a Rabi coupling between components 
$|1,0\rangle$ and $|1,+1\rangle$ for a finite time interval, atoms are transferred
to $\psi_{+1}$ and a bright soliton is formed there too. After switching off
the Rabi coupling between $\psi_0$ and $\psi_{+1}$, the percentage population
of atoms in the three components is $97:2:1$.


We have also performed numerical simulations for the DBB solitons in the presence 
of a trap, in the case where the location of the soliton is displaced from the trap center. 
We have confirmed in such a case that the DBB solitons generically
perform robust oscillations inside the trap.
A typical example is shown in Fig. \ref{osc} illustrating that --despite the potential presence
of sound waves inside the condensate-- the oscillation
persists for very long times of the order of many seconds.

Apart from DBB solitons, in our experiments we have also observed the emergence of 
DDB ones, again with lifetimes on the order of hundreds of milliseconds. 
To generate DDB solitons,
a procedure similar to that of the DBB soliton generation is followed. 
We begin with all atoms in the $|1,0\rangle$ state. A small fraction of atoms
is then transferred from the $|1,0\rangle$ state to the $|1,+1\rangle$ state. Subsequently,
a weak magnetic gradient is again applied and leads to the formation of DB solitons. 
In this experiment, the dark solitons reside in the $|1,0\rangle$ state, while 
the bright soliton components are formed by atoms in the 
$|1,+1\rangle$ state. After the DB solitons are formed, the magnetic gradient
is removed, which is necessary to ensure long lifetimes of the solitonic structures.
To convert the DB solitons into DDB ones, an RF sweep is used to transfer
an adjustable fraction of the atoms from the $|1,0\rangle$ state to the $|1,-1\rangle$ state.
This completes the formation of a DDB soliton.  In our experiments, we have found that the DDB solitons 
(and also the DBB solitons discussed above) have lifetimes on the order 
of hundreds of milliseconds.

\begin{figure}[tbp]
\begin{center}
\includegraphics[width=3.4in]{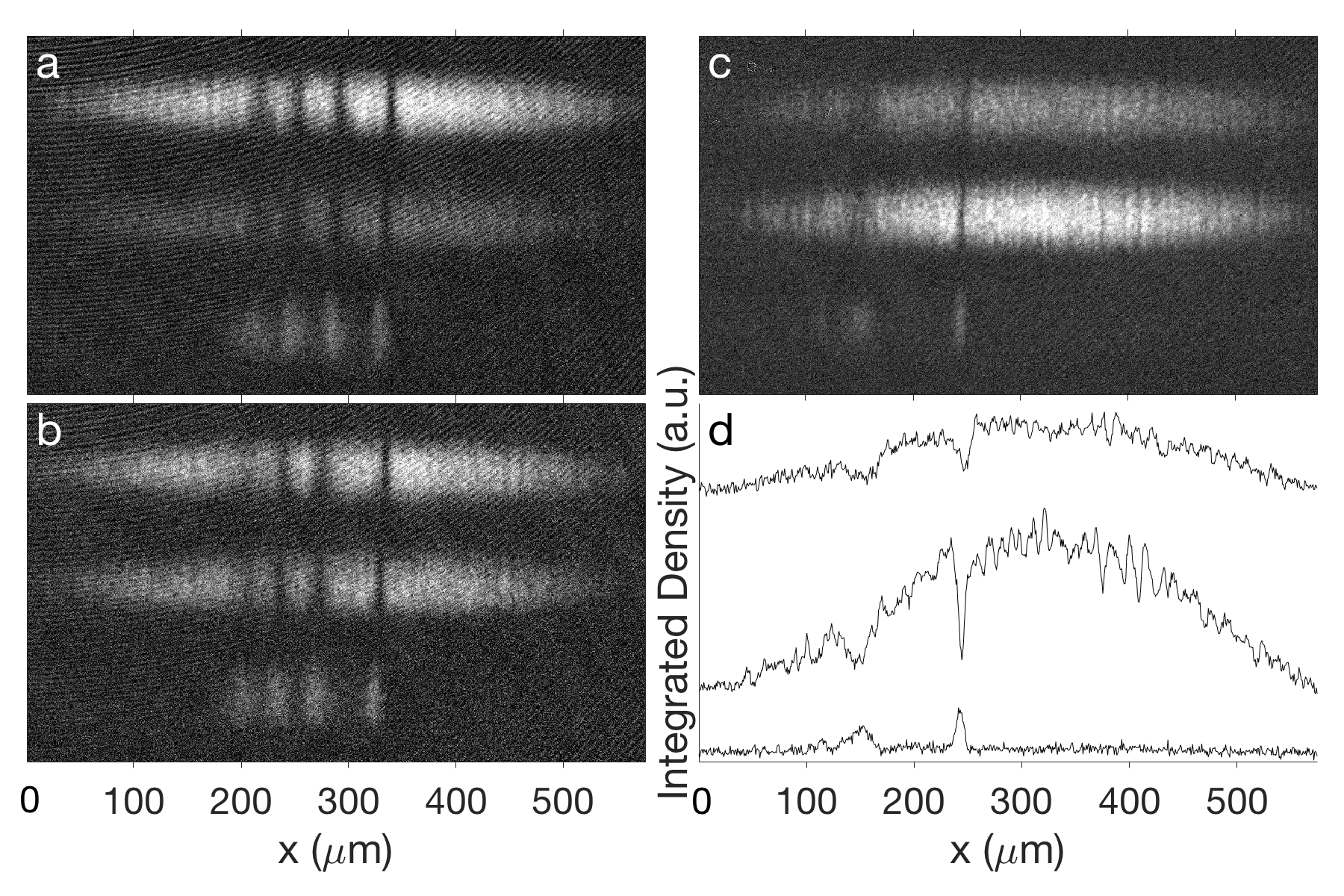}
\caption{a) - c) Experimental ToF images of DDB solitons. Here it is shown that the DDB solitons can be generated for a large variation of the relative populations of the $|1,-1\rangle$ , $|1,0\rangle$ , and $|1,+1\rangle$ states (upper, middle, and lower cloud in each image, respectively). In each case, to verify the stability of the soliton a $100~\text{ms}$ in-trap evolution time is applied after DDB soliton formation.
The relative populations of the three states are
a) $71:21:8$ ,  b) $53:38:9$ , and c) $33:66:1$. d) Integrated density profiles of the Zeeman levels of image (c). The plots are offset in the y-dimension for clarity and to mimic the spatial order of each state in the ToF image.}
\label{DDB_diff_comp}
\end{center}
\end{figure}

The existence of these features appears to be fairly insensitive to the exact population
ratio of the three Zeeman states. For example, we have experimentally verified
the existence of DDB structures for different percentage population of atoms in the
three states including  $71:21:8$ , $53:38:9$ , and $33:66:1$. These results highlight the generic robustness of the DDB structures. A pertinent example is shown in Fig.~\ref{DDB_diff_comp}.

The formation of DDB solitons can also be predicted in the framework of the 
multiscale expansion method. In this case, assuming approximately equal chemical potential 
for all spin components, $\mu \approx [\lambda_s(1+r^2)-\lambda_a (1-r^2)]n_0$  
(where $n_0$ is the pertinent steady-state density and $r=|\psi_{0}|/|\psi_{-1}|$), we can show 
that DDB solitons do exist, and assume the following form:
\begin{equation}
\begin{array}{rcl}
\!\!\!\!\!\!\!\!\!\!\!\!
\psi_{-1}&\approx&\sqrt{n_0 +\epsilon \rho(X,T)}
e^{-i\mu t + i\epsilon^{1/2}\sqrt{\mu}\int \rho(X,T) dX},
\\[1.0ex]
\psi_{0}&=& r\psi_{-1}, \quad
\psi_{+1}\approx \epsilon^{3/4} q(X,T)e^{i[\sqrt{\mu}x-(3/2)\mu t]},
\label{ps1}
\end{array}
\end{equation}
where $\epsilon$ is again a small parameter, 
$X=\epsilon^{1/2}(x-\sqrt{\mu}t)$ and $T=\epsilon t$ are stretched variables, 
and the functions $\rho(X,T)$ and $q(X,T)$ are governed by the equations:
\begin{equation}
\begin{array}{rcl}
&&\partial_{T}\rho =-\sqrt{\mu}(\lambda_s+\lambda_a)\partial_{X}\left( |q|^{2}\right),
\\[1.0ex]
&&i\partial_{T}q+\frac{1}{2}\partial _{X}^{2}q-(\mu/n_0)\rho q=0.
\label{yo}
\end{array}
\end{equation}
The above equations constitute the single component Yajima-Oikawa (YO) system~\cite{yajima}.
The YO system is completely integrable and possesses soliton solutions of the form
$\rho \propto -{\rm sech}^2(k_s X-\omega_s T)$ and $q \propto {\rm sech}(k_s X-\omega_s T)$,
where $k_s,~\omega_s$ are constants. These expressions, when
substituted into Eqs.~(\ref{ps1}), give rise to approximate
DDB solitons, for the $m_{F}=-1,0,+1$ spin components, respectively.
Note that we have found (results not shown here) that such DDB solitons 
are also long-lived in our direct numerical simulations.

In conclusion, we have demonstrated the creation of 
dark-bright-bright (DBB) and dark-dark-bright (DDB) solitons in a spinor $F=1$ $^{87}$Rb condensate. 
It was found that these structures are quite robust, featuring lifetimes on the order
of several hundreds of milliseconds, and can be formed for different relative populations
of atoms in the three Zeeman states. We have employed a perturbative approach to show
that these mixed solitons can be approximated by solutions of the multi- and single-component  
Yajima-Oikawa systems. This connection also underscores the breadth of 
relevance of these patterns and supports their robustness. 
Direct numerical simulations corroborate our results indicating that these solitons 
can persist but also that they can robustly oscillate inside the condensates.

The experimental, theoretical and numerical manifestation
of such states paves the way for a number of interesting studies in
the future. For instance, it will be particularly relevant to
explore more systematically the oscillations of these solitary
waves in the trap and to identify their oscillation frequency 
as a function of both the trap frequency and the atomic fractions
of the different components, similarly to 
the cases of one- and two-components~\cite{busch1,busch2}.
Another possibility is to explore the generalizations of such
spinorial states in higher dimensions constructing spinorial
analogues of vortex-bright (or baby-skyrmion, or filled-core
vortex) states~\cite{emergent} and 
understand their dynamics and
interactions. Soliton interaction dynamics and stability over parametric
variations (e.g., of the spin-dependent part of the Hamiltonian)
would also be particularly relevant to consider even in the one-dimensional
case. More broadly, spinor BECs open an avenue
to proceed beyond 2-component soliton dynamics that we expect will
yield exciting developments in the near future.


\begin{thebibliography}{99}
\bibitem{imaging} For imaging, the trap is suddenly switched off and a vertical magnetic gradient is applied. Due to this gradient, the three Zeeman states are separated along the vertical z-direction after a brief time of flight (13~ms). An image is then taken along the y-direction, showing the position of the three atom clouds.
\end{thebibliography}
\end{document}